\documentclass{article}
\usepackage{spconf}
\usepackage{times}
\usepackage{soul}
\usepackage{url}
\usepackage{xspace}
\usepackage{siunitx}
\usepackage[hidelinks]{hyperref}
\usepackage[utf8]{inputenc}
\usepackage[small]{caption}
\usepackage{graphicx}
\usepackage{amsmath}
\usepackage{amsthm}
\usepackage{booktabs}
\usepackage{algorithm}
\usepackage{algorithmic}
\usepackage{multirow}
\usepackage[switch]{lineno}
\usepackage{cleveref}
\usepackage{tabularx}
\usepackage{array}
\usepackage{subcaption}
\usepackage{enumitem}
\usepackage{pifont}
\usepackage{colortbl} 
\usepackage{bibspacing}
\usepackage{lipsum}
\newcolumntype{C}{>{\centering\arraybackslash}X}
\crefname{section}{§}{§§}
\Crefname{section}{§}{§§}
\ninept

\makeatletter
\renewcommand{\paragraph}[1]{%
  \par\vspace{0.3em}
  \noindent\textbf{#1}\quad
}
\makeatother

\title{Why Do Speech Language Models Fail to Generate Semantically Coherent Outputs? A Modality Evolving Perspective}
\name{Hankun Wang, Haoran Wang, Yiwei Guo, Zhihan Li, Chenpeng Du,  $^{\dagger}$Kai Yu \thanks{$^{\dagger}$\scriptsize{Kai Yu is the corresponding author.}}}
\address{
X-LANCE Lab, School of Computer Science, Shanghai Jiao Tong University, China \\
MoE Key Lab of Artificial Intelligence, Jiangsu Key Lab of Language Computing, China \\
\texttt{\{wanghankun, kai.yu\}@sjtu.edu.cn}}
\begin{document}
\maketitle
%
\begin{abstract}
Although text-based large language models exhibit human-level writing ability, end-to-end speech language models (SLMs) still struggle to generate semantically coherent outputs without explicit text transcription. There are several potential reasons for this performance degradation: (A) speech tokens mainly provide phonetic information rather than semantic information, (B) the length of speech sequences is much longer than that of text sequences, and (C) paralinguistic information, such as prosody and accent, introduces additional variability. In this paper, we explore the influence of three key factors separately by transitioning the modality from text to speech in an evolving manner. Our findings reveal that the impact of the three factors varies. Factor A has a relatively minor impact, factor B influences syntactical and semantic modeling more significantly, and factor C exerts the most substantial impact, particularly in basic lexical modeling. Based on these findings, we provide insights into the unique challenges of training SLMs and highlight pathways to develop more effective end-to-end SLMs.
\end{abstract}
\vspace{-1pt}
\section{Introduction}
\label{sec:intro}
Constructing end-to-end speech generation models is one of the ultimate goals in the field of speech.
Despite the proven effectiveness of auto-regressive (AR) \emph{text} Large Language Models~\cite{brown_language_2020,dubey_llama_2023}, building a \emph{Speech} Language Model (SLM) that can generate semantically coherent speech \textit{without text transcription guidance} is still an open problem.
Recently, the mainstream solution for speech generation systems is to rely on transcription guidance~\cite{cui_recent_2024,ji_wavchat_2024}. 
Multiple works~\cite{xie_mini-omni_2024,defossez_moshi_2024,chen2025minmo,xu2025qwen2} have adopted a two-step approach. First, an LLM processes the input audio and instructions to generate a textual response. Then, the text output serves as an explicit guide during speech synthesis. 
This approach leverages the mature abilities of text LLMs and text-to-speech (TTS) models, enabling more stable and semantically coherent generation. However, several factors limit the performance ceiling of such architectures. For instance, the TTS model in this architecture lacks an understanding of paralinguistic and other non-textual information present in the input. It also struggles to generate highly natural non-verbal segments. Additionally, the wealth of in-the-wild speech data cannot be fully utilized for training. Therefore, exploring a truly end-to-end speech generation model without transcription guidance is essential and urgently demanded. 

\begin{table}[t]
\centering
\caption{Overview of generalized speech language modeling tasks. In \textit{Input} and \textit{Output} columns, \textit{T} is for text and \textit{S} is for speech. The \textit{Trans. Guid.} column indicates whether text transcription guidance is used for synthesizing speech. This paper focuses on investigating the challenges associated with the last row.}
\vspace{-7pt}
\scalebox{0.8}{
\begin{tabular}{lcccl}
\hline
\textbf{Speech Task}                 & \textbf{Input} & \textbf{Output} & \textbf{Trans. Guid.} & \textbf{Representative Work} \\ \hline \hline
Text-to-Speech                & T           & S          &   \ding{51}             &     VALL-E~\cite{wang_neural_2023}      \\
Understanding & S         & T            &           -   &                           SALMONN~\cite{tang_salmonn_2024} \\
Interaction            & S         & T + S          &      \ding{51}        &          Qwen2.5-Omni~\cite{xu2025qwen2} \\
\rowcolor[gray]{0.9} Language Model      & S         & S          &   \ding{55}           &  GSLM~\cite{lakhotia_generative_2021} \\ \hline
\end{tabular}}
\label{tab:slm-disamb}
\vspace{-12pt}
\end{table}

We follow definitions in Table~\ref{tab:slm-disamb}, where \textit{SLM} refers to models that generate speech without text guidance. Since GSLM~\cite{lakhotia_generative_2021} and AudioLM~\cite{borsos_audiolm_2023}, transformer-based SLMs still trail behind text-guided systems. Prior work has attempted lowering frame rates~\cite{lakhotia_generative_2021,hassid_textually_2024,cho_sylber_2024,cuervo_scaling_2024} or aligning speech with text~\cite{zhang_speechgpt_2023,nguyen_spirit-lm_2024,hassid_textually_2024}, but none fully resolve the coherence gap. Meanwhile, the underlying reasons for their limitations remain unexplored. As a result, the community lacks insight into the differences between how SLMs and text LLMs work, and current improvements in SLMs are largely empirical attempts to approximate text LLMs in terms of data length and form.

In this paper, we systematically analyze the low performance of SLMs based on discrete semantic speech tokens and aim to answer the fundamental question below:
\paragraph{Question} \emph{What are the critical factors that make the speech modality significantly harder to train compared to the text modality?}
Possible factors are: 
\begin{itemize}[itemsep=1pt,topsep=2pt,left=2pt]
    \item \textbf{(A)} Speech tokens such as HuBERT are more phonetic than truly semantic~\cite{choi_self-supervised_2024}. Extracting semantic information becomes harder when using phonetic-based representations.
    \item \textbf{(B)} The length of the speech token sequence is considerably longer than its transcription text token sequence since the pronunciation duration information is included in the speech sequence by assigning each frame a token.
    \item \textbf{(C)} The sequence retains some paralinguistic information, such as prosody and timbre, introducing additional variability.
\end{itemize}


To answer this question, we propose viewing the significant gap between text and speech modalities from an evolving perspective (\cref{sec:text-speech-evo}). 
We train separate LMs on the same speech dataset, using different modalities: text-based, phone-based, and speech-based token representations. The differences between modalities correspond to the possible factors listed in the question. Therefore, by evaluating the performance of LMs trained by these modalities in various tasks (\cref{sec:exp-setup}), a systematical study is established and the impact of these factors can be comprehensively analyzed. 

Our findings reveal that the three factors affect performance to varying degrees. Factor A has a relatively minor impact, factor B more noticeably influences syntactic and semantic modeling, and factor C exerts the most significant impact, particularly in lexical modeling (\cref{sec:results}).
Based on the experimental findings, we propose a few possible ways to achieve end-to-end speech modeling (\cref{sec:future-dir}).

\section{Related Works}
We categorize prior efforts on SLMs mainly into two directions: reducing representation bit rates and aligning speech with text.

\paragraph{Reducing Bit Rates} 
GLSM~\cite{lakhotia_generative_2021} proposed that lower frame-rate, self-supervised semantic representations facilitate language modeling, which uses de-duplicated HuBERT~\cite{hsu_hubert_2021} to achieve an average frame rate below \SI{40}{\hertz}. Other works create their own semantic speech tokens with lower frame rates, reaching \SI{25}{\hertz}~\cite{hassid_textually_2024}, \SI{20}{\hertz}~\cite{shen_acoustic_2024} and even \SI{5}{\hertz}~\cite{baade_syllablelm_2024,cho_sylber_2024,tseng2025taste}. 
However, this approach faces a hard trade-off between preserving semantic clarity and scalability while maintaining audio reconstruction quality.

\paragraph{Aligning with Text}
The second direction is to align speech with text in terms of representations, model architecture, model parameters, or training schemes.
TWIST~\cite{hassid_textually_2024} finds that initializing SLM with a pre-trained text LLM can enhance its performance. The SpeechGPT works~\cite{zhang_speechgpt_2023} utilize speech-text-paired data for the model fine-tuning process. SpiritLM~\cite{nguyen_spirit-lm_2024} interleaves speech and text tokens at the word level during pre-training. 
Align-SLM~\cite{lin2024align} uses ASR transcription to build a reinforcement learning curriculum with LLM feedback. Other methods include employing novel model architectures used in text LMs~\cite{park2024longform} and group-wise generation~\cite{zhang2024intrinsicvoice}.
Although this modality alignment approach improves SLMs to some extent, the models still struggle to synthesize semantically coherent speech without text guidance.

\vspace{-5pt}
\section{Modality Evolving}
\label{sec:text-speech-evo}
\vspace{-7pt}

\begin{table}[t]
\caption{Modalities overview. The \textit{Vocab Size} column shows the vocabulary size of the modality. The \textit{\#Tokens} column represents the number of encoded tokens of the training set. The \textit{\#Tokens/s} column represents the average number of tokens per second. The \textit{Factor} column represents the corresponding factor ID explored by the modality.}
\vspace{-8pt}
\centering
\scalebox{0.83}{
\begin{tabular}{l|cccc}
\toprule
\textbf{Modality}               & \textbf{Vocab Size} & \textbf{\#Tokens} & \textbf{\#Tokens/s} & \textbf{Factor} \\ \midrule
\texttt{Text-BPE}    & 2048              &      696.2M    & 4.45 & \multirow{2}{*}{Topline} \\
\texttt{Text-Raw}     & $\sim$80         &      2.249B    & 14.53 & \\ \hline
\texttt{Phone-BPE}   & 2048              &      625.7M    & 4.04 & \multirow{2}{*}{+A}     \\ 
\texttt{Phone-Raw}    & $\sim$80         &      1.542B    & 9.97 \\ \hline
\texttt{Phone-Repeat} & $\sim$80         &      7.737B    & 50 & +B               \\ \hline
\texttt{Speech-HuBERT} & 2048            &      7.737B    & 50 & +C               \\ \bottomrule
\end{tabular}}
\label{tab:modal-overview}
\vspace{-10pt}
\end{table}

\begin{table*}[t]
\caption{Main results: impact of three factors on task performance. Relative changes in accuracy ($\Delta$Acc\%) on sWUGGY, sBLIMP, and Topic-SC, and relative changes in perplexity ($\Delta$PPL\%) on the continuation task are reported.}
\vspace*{-5pt}
\centering
\label{tab:infl-factor-cvg}
\scalebox{0.85}{
\begin{tabular}{lcl|cccc}
\hline 
\multirow{2}{*}{\textbf{Baseline Modality}} & \multirow{2}{*}{\textbf{Factor}} & \multirow{2}{*}{\textbf{Resulting Modality}} & \textbf{sWUGGY} & \textbf{sBLIMP} & \textbf{Topic-SC} & \textbf{Continuation} \\ 
                           &                 &                             & $\Delta$Acc\% & $\Delta$Acc\% & $\Delta$Acc\%   & $\Delta$PPL\%        \\ \hline \hline
\texttt{Text-BPE}          & +A              & \texttt{Phone-BPE}          & -0.0                          & +0.0                           & -3.7                              & +7.8        \\
\texttt{Text-Raw}          & +A              & \texttt{Phone-Raw}          & +0.0                          & +1.6                           & +0.9                              & +26.6        \\
\texttt{Phone-Raw}         & +B              & \texttt{Phone-Repeat}       & -0.3          & -11.1          & -12.5            & +88.3               \\ 
\texttt{Phone-Repeat}      & +C              & \texttt{Speech-HuBERT}      & -40.6         & -13.4          & -9.3             & +140.7              \\
\hline
\end{tabular}}
\vspace{-2pt}
\end{table*}

\begin{figure*}[t]
    \centering
    \begin{subfigure}{0.23\textwidth}
        \includegraphics[width=\textwidth]{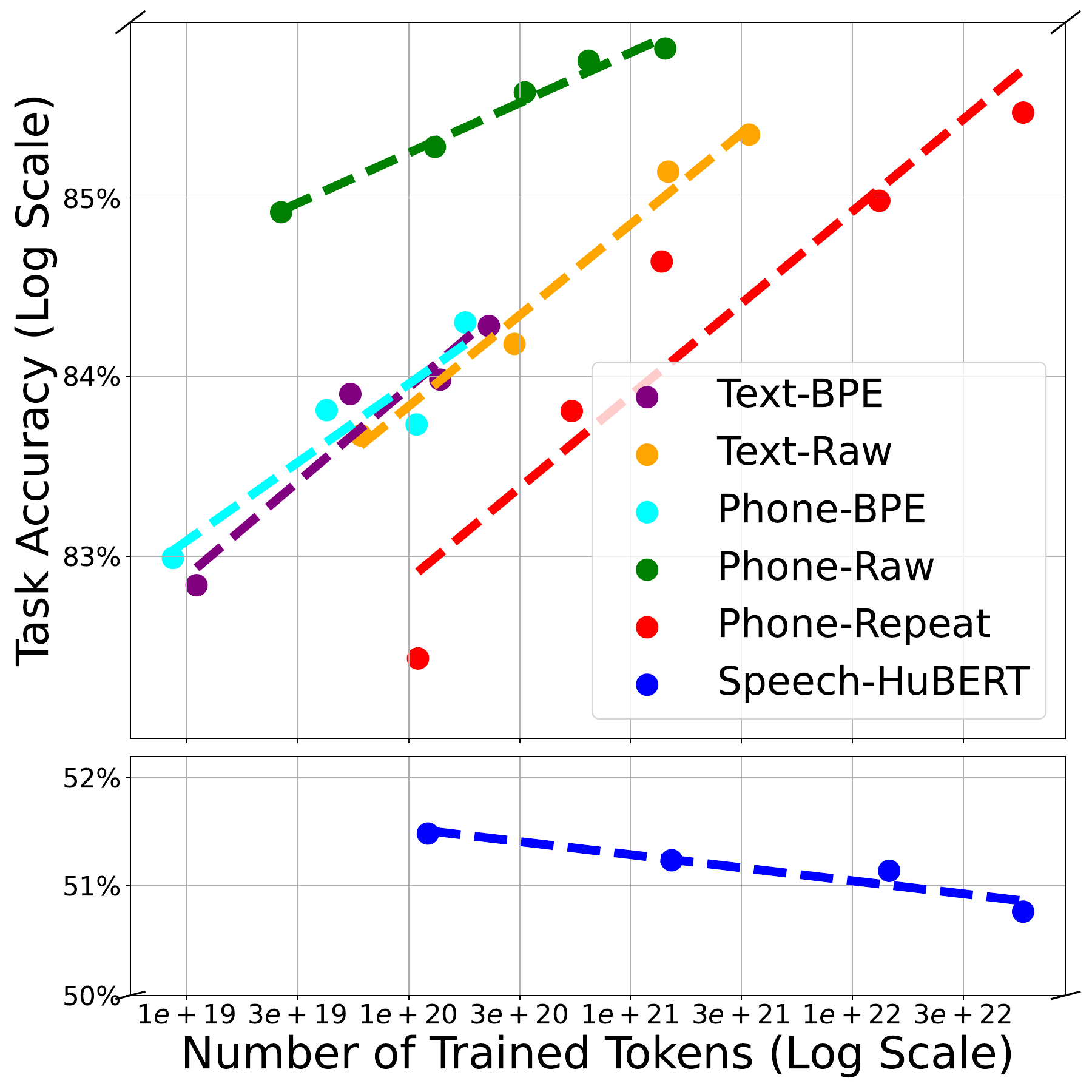}
        \caption{sWUGGY (Lexical)}
    \end{subfigure}
    \hspace*{25pt}
    \begin{subfigure}{0.23\textwidth}
        \includegraphics[width=\textwidth]{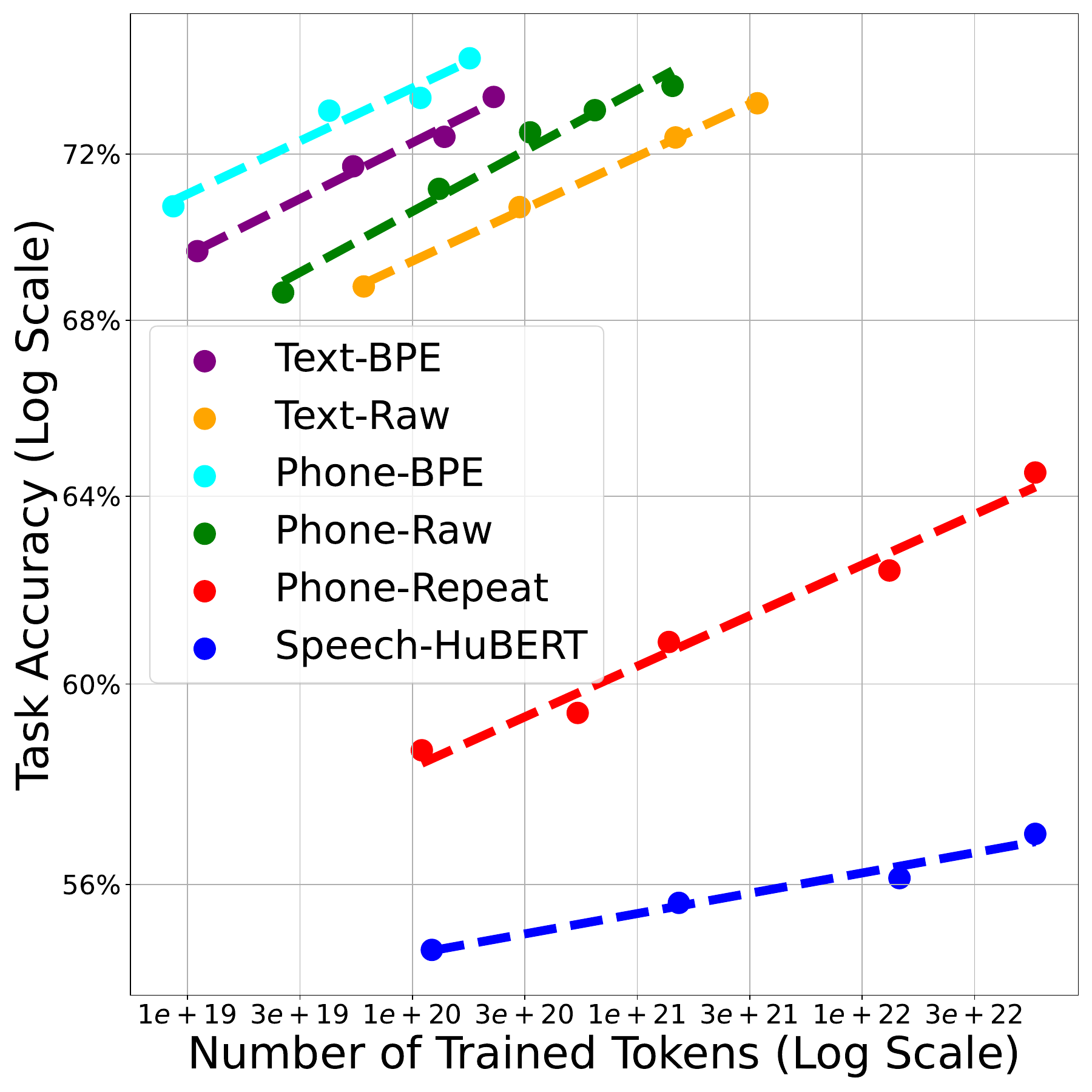}
        \caption{sBLIMP (Syntactic)}
    \end{subfigure}
    \hspace*{25pt}
    \begin{subfigure}{0.23\textwidth}
        \includegraphics[width=\textwidth]{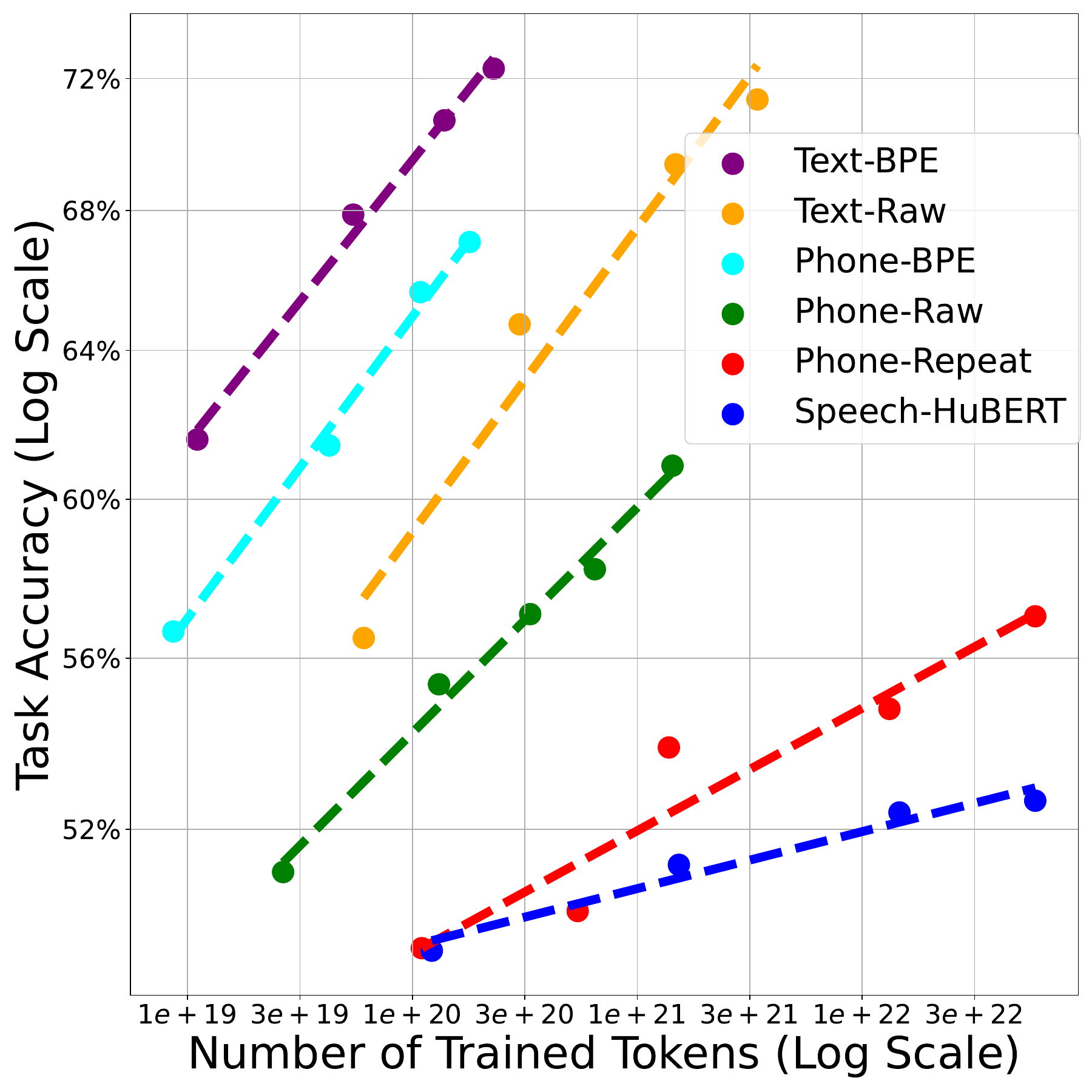}
        \caption{Topic-StoryCloze (Semantic)}
    \end{subfigure}
\vspace{-8pt}
    \caption{Results after training the same number of tokens (within the first epoch).}
    \label{fig:main-res-scale}
\vspace{-9pt}
\end{figure*}

\begin{table}[th]
\centering
\caption{Absolute accuracies (\%) on the objective tasks and perplexities on the continuation task. All LMs are trained to convergence.}
\vspace{-5pt}
\label{tab:main-res-cvg}
\scalebox{0.82}{
\begin{tabular}{l|ccccc}
\toprule 
\multirow{3}{*}{\textbf{Modality}} & \textbf{sWUGGY} & \textbf{sBLIMP}        & \textbf{Topic-SC} & \multicolumn{2}{c}{\textbf{Continuation}}        \\ \cline{2-6} 
& \multirow{2}{*}{\textbf{Acc.}(\%)$\uparrow$} & 
 \multirow{2}{*}{\textbf{Acc.}(\%)$\uparrow$} & 
 \multirow{2}{*}{\textbf{Acc.}(\%)$\uparrow$} & 
 \multicolumn{2}{c}{\textbf{PPL}$\downarrow$}  
 \\ 
& & & & mean & std 
\\ \midrule 
\texttt{Text-BPE}             &          85.1        &        74.9          &  \textbf{73.6}          &      \textbf{51.3} & 32.0                               \\
\texttt{Text-Raw}              &          85.6        &        73.3          &           66.0          &      54.6 & 33.4                              \\ \hline
\texttt{Phone-BPE}               &          85.0        &\textbf{75.0}         &           70.9          &   59.1 & 42.9                                     \\
\texttt{Phone-Raw}                & \textbf{85.8}        &        74.5          &           66.6          &   69.1 & 58.9                               \\ \hline
\texttt{Phone-Repeat}             &          85.5        &        66.2          &           58.3          &   130.1 & 283.6                             \\ \hline
\texttt{Speech-HuBERT}             &          50.8        &        57.3          &           52.9          &   313.2 & 296.1                              \\ 
\bottomrule
\end{tabular}}
\vspace{-6pt}
\end{table}

\subsection{Overview}
This section introduces the modalities used in our study, which progressively evolve from text to phones and then to speech. This perspective allows us to pinpoint where the shift in modality leads to significant performance degradation. Table~\ref{tab:modal-overview} provides a summary.

\paragraph{Text-Based Modalities}
We use two text modalities with different tokenization strategies:
\begin{itemize}[topsep=3pt,leftmargin=10pt,parsep=1pt,itemsep=1pt]
    \item \textbf{Text-BPE}: A subword-based tokenizer with 2048 units, trained using SentencePiece~\cite{kudo_sentencepiece_2018} on LibriHeavy-medium~\cite{kang_libriheavy_2024} transcriptions. It serves as the topline semantic representation, close to standard LLM tokenizers.  
    \item \textbf{Text-Raw}: A character-level tokenizer (letters, digits, punctuation), which provides a simple baseline for comparison with phone-level units.  
\end{itemize}

\paragraph{Phone-Based Modalities}
Phones act as the bridge between text and speech. We study three phone-level variants:
\begin{itemize}[topsep=3pt,leftmargin=10pt,parsep=1pt,itemsep=1pt]
    \item \textbf{Phone-Raw}: Each phone is a token ($\sim$80 types of phones, including silence). Sequences are derived from Kaldi alignments to retain pronunciation information.  
    \item \textbf{Phone-BPE}: Built on \texttt{Phone-Raw} with a BPE tokenizer (same vocab size as \texttt{Text-BPE}). This enables a fair comparison of phonetic vs. semantic subword units.  
    \item \textbf{Phone-Repeat}: Phone tokens repeated according to duration, resampled to \SI{50}{Hz}, aligning with speech token frame rates. This tests the effect of sequence length.  
\end{itemize}

\paragraph{Speech-Based Modality}
Numerous discrete speech representations have been explored in prior research~\cite{hsu_hubert_2021,baevski_wav2vec_2020,defossez_high_2022,kumar_high-fidelity_2023,ju_naturalspeech_2024}. In this work, we adopt discrete HuBERT representations as the target for LMs. This choice aims to add a modest amount of paralinguistic information while preserving the rich phonetic content~\cite{guo2025recent}.
\begin{itemize}[topsep=3pt,leftmargin=10pt,parsep=1pt,itemsep=1pt]
    \item \textbf{Speech-HuBERT}: Discrete tokens obtained by clustering HuBERT-Large hidden states into 2048 units at \SI{50}{Hz}. Compared to phones, these tokens add paralinguistic information while preserving phonetic content.
\end{itemize}

\section{Experimental Setup}
\label{sec:exp-setup}

\paragraph{Datasets}
We use LibriHeavy-large \cite{kang_libriheavy_2024} (a filtered subset of LibriLight-60k \cite{kahn_libri-light_2020}, resulting in $\sim$50k hours of speech) as training data. Text transcriptions are filtered to English characters only. Phone-level data is obtained using Kaldi. Speech tokens are extracted with the HuBERT-large checkpoint.

\paragraph{Hyperparameters}
All LMs adopt TinyLlama~\cite{zhang_tinyllama_2024} (22 Transformer layers, 32 heads, Group Query Attention~\cite{ainslie_gqa_2023}, 1.1B params), trained from scratch with AdamW~\cite{loshchilov_decoupled_2019}, learning rate \num{4e-4}, cosine scheduler. Training uses 4$\times$NVIDIA-A800-80G, global batch size 128, with per-batch utterances padded to the maximum window length. Models are trained to convergence based on validation loss.

\paragraph{Tasks}
Evaluation is performed in a zero-shot setting on three objective discriminative tasks and one continuation task. The test data of three objective discriminative tasks are transformed into modalities listed in Table~\ref{tab:modal-overview} for the evaluation of the corresponding LM.
The four tasks are:
\begin{itemize}[topsep=3pt,leftmargin=10pt,parsep=1pt,itemsep=1pt]
    \item \textbf{sWUGGY}~\cite{nguyen_zero_2020}, which evaluates lexical modeling abilities. Each sample is a word pair (real vs.~pseudo) provided in speech, text, and phone forms. The LM computes likelihoods for both candidates; success if the real word receives higher likelihood.
    
    \item \textbf{sBLIMP}~\cite{warstadt_blimp_2020,nguyen_zero_2020}, which evaluates syntactic modeling abilities. Each sample is a sentence pair (grammatical vs.~ungrammatical). Data is available in speech and text, with phone sequences obtained via Kaldi alignments. The LM is correct if it assigns higher likelihood to the grammatical sentence.
    
    \item \textbf{Topic-StoryCloze (Topic-SC)}~\cite{hassid_textually_2024,mostafazadeh_lsdsem_2017}, which evaluates semantic modeling abilities. Each instance consists of a short base story and two candidate continuations. The LM selects the more plausible continuation by comparing likelihoods.
    
    \item \textbf{Continuation task}, which is free autoregressive generation. We design 20 prompts of varying lengths and content, 
     transformed into each modality. Decoding uses temperature $\in[1.0,1.2]$, top-p=0.9, with 10 generations per prompt using different seeds. Outputs from non-text modalities are transcribed into text by using Whisper-large-v3~\cite{radford_2023_robust_whisper} model, and perplexity is computed with pretrained Llama-3.1-8B
     ~\cite{dubey_llama_2023}. More details can be found at \href{https://x-lance.github.io/SLM-evolving/}{\small\texttt{https://x-lance.github.io/SLM-evolving/}}. 
    
\end{itemize}

\section{Results and Analysis}
\label{sec:results}


\begin{figure*}[t]
    \centering
    \begin{subfigure}{0.32\textwidth}
        \includegraphics[width=\textwidth]{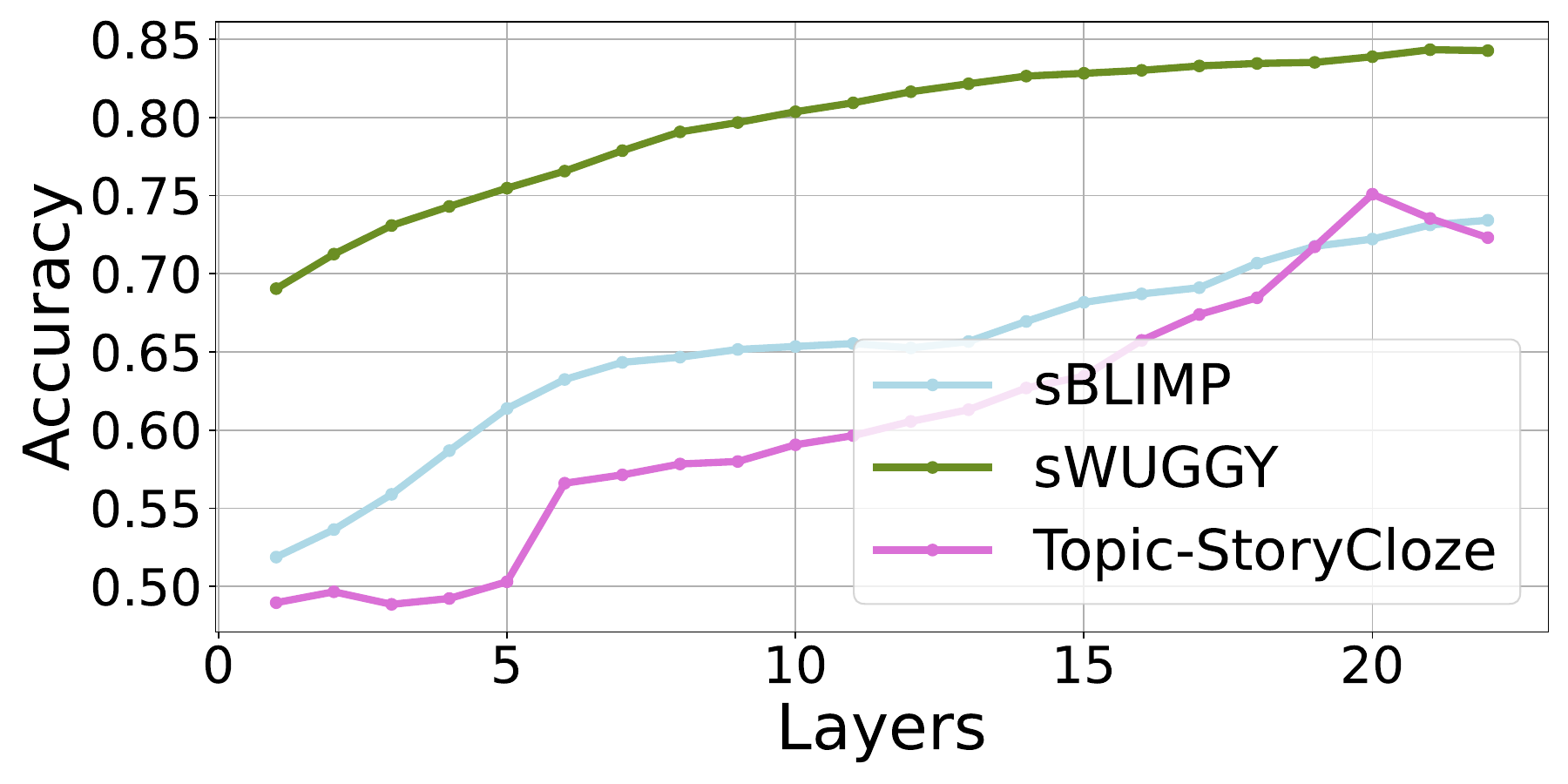}
        \vspace*{-15pt}
        \caption{\texttt{Text-BPE} Modality}
        \label{fig:text-bpe-by-layers}
    \end{subfigure}\hspace{6pt}
    \begin{subfigure}{0.32\textwidth}
        \includegraphics[width=\textwidth]{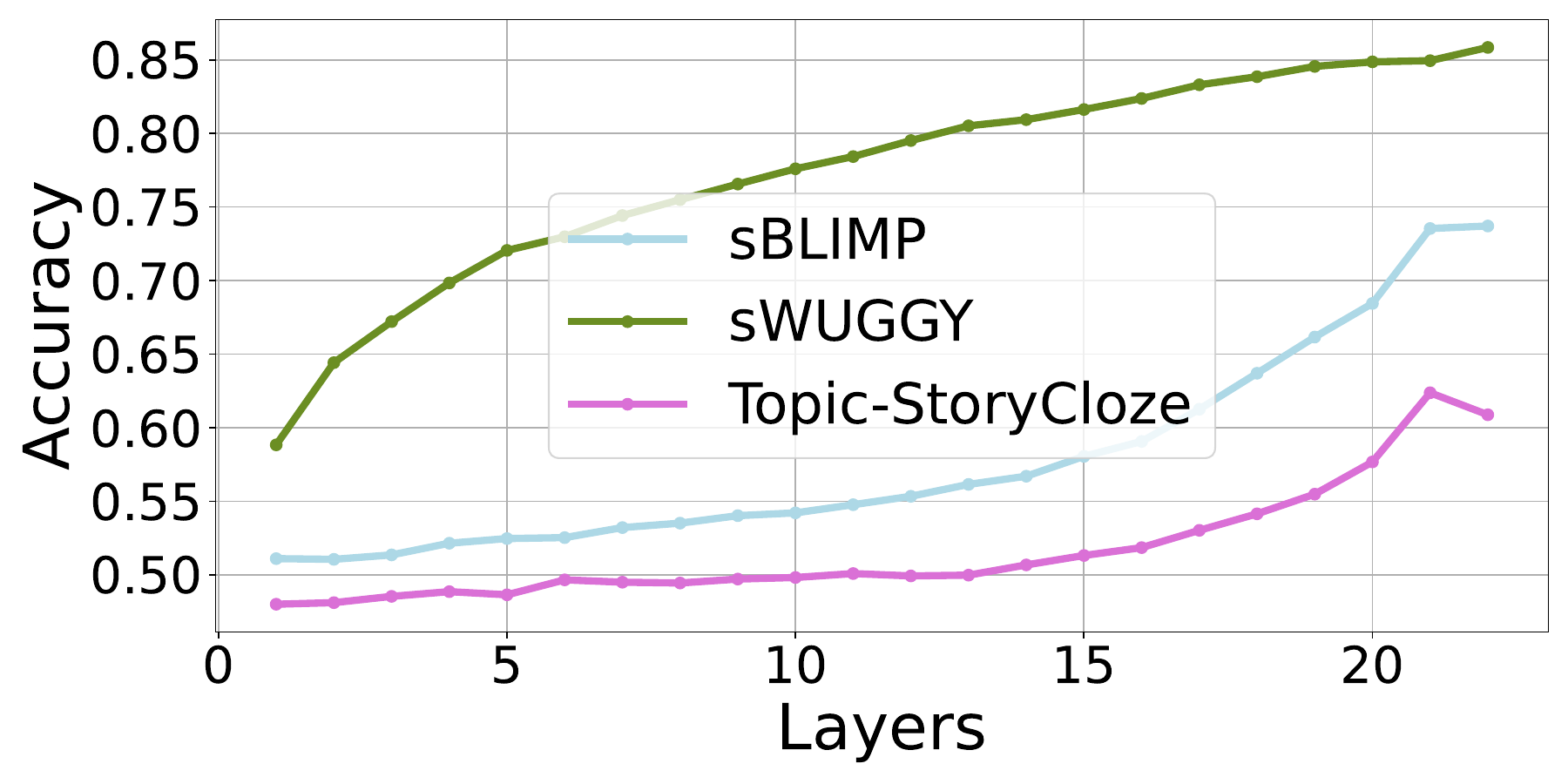}
        \vspace*{-15pt}
        \caption{\texttt{Phone-Raw} Modality}
        \label{fig:phone-raw}
    \end{subfigure}\hspace{6pt}
    \begin{subfigure}{0.32\textwidth}
        \includegraphics[width=\textwidth]{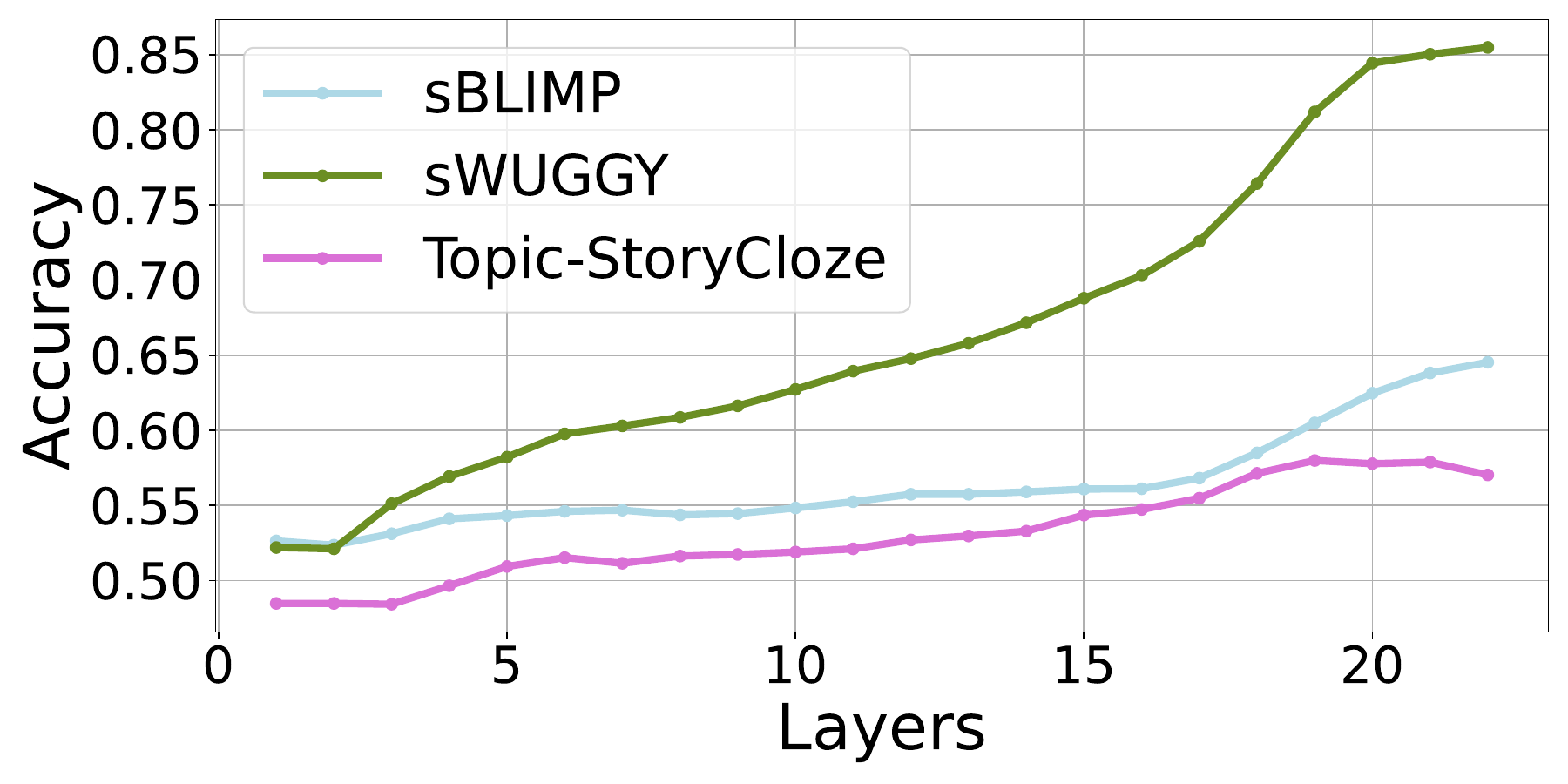}
        \vspace*{-15pt}
        \caption{\texttt{Phone-Repeat} Modality}
        \label{fig:phone-repeat-by-layers}
    \end{subfigure}
    \vspace*{-6pt}
    \caption{Accuracy results of internal layers outputs for all objective tasks. 
    }
    \vspace*{-16pt}
    \label{fig:internal-out-ana}
\end{figure*}
\begin{figure}[t]
    \centering
    \includegraphics[width=0.4\textwidth]{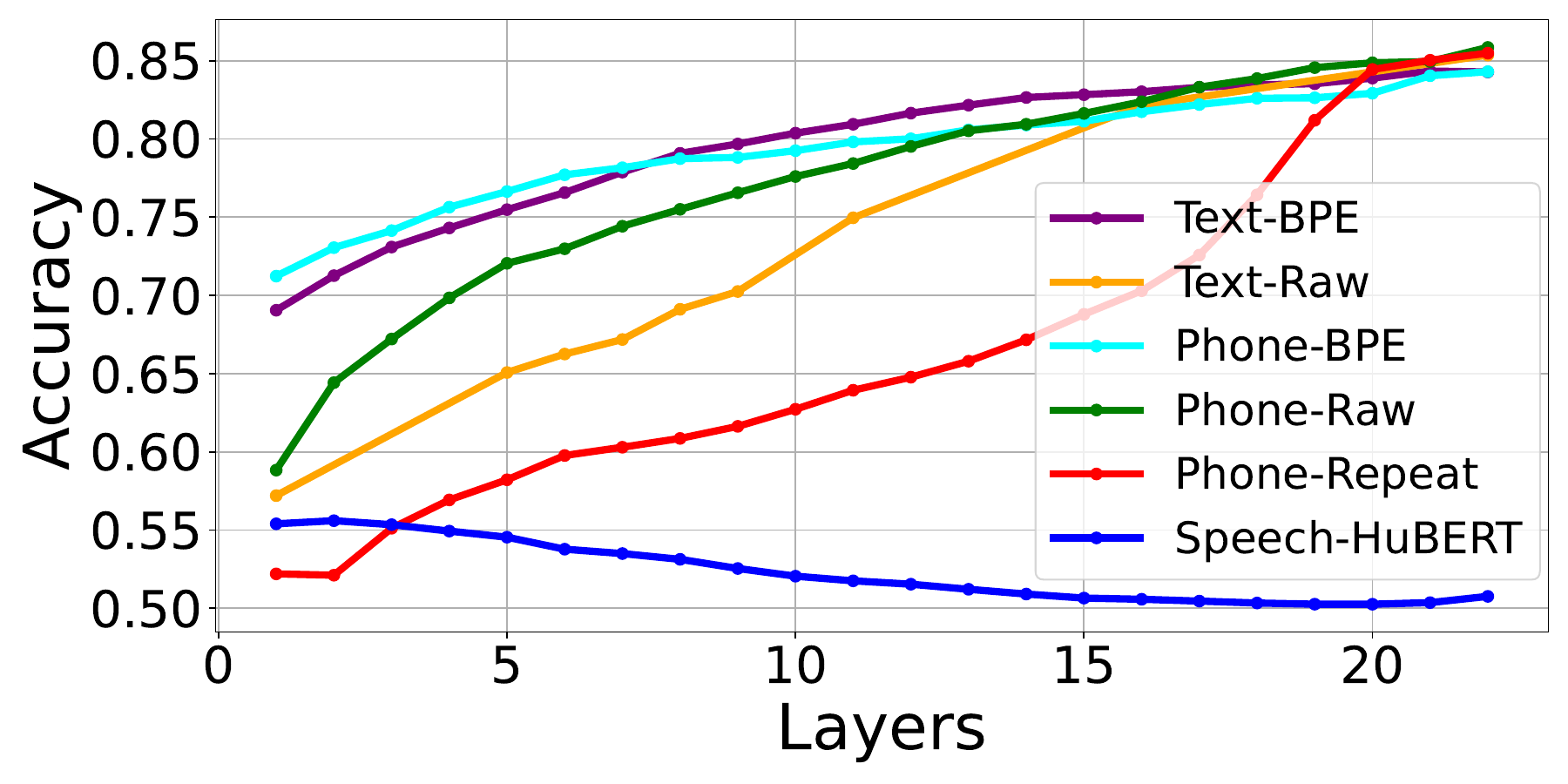}
\vspace{-6pt}
    \caption{Layer-wise accuracy changes for the sWUGGY task.}
    \label{fig:swuggy-by-layers}
\vspace{-6pt}
\end{figure}
\subsection{Comparison: LMs of Different Modalities}
We first compare the LMs when they have learned the same amount of semantic information, so the LMs are trained on the same dataset in their respective modalities until the validation loss converges. The results of three objective tasks and the continuation task are shown in Table~\ref{tab:main-res-cvg} and Table~\ref{tab:infl-factor-cvg}. 

For the lexical task, text-based and phone-based modalities achieve similar high accuracy, exceeding 85\%, implying that factors A and B have a minor impact on lexical modeling. In contrast, the \texttt{Speech-HuBERT} modality performs only slightly better than the random baseline of 50\%. This highlights the substantial difficulty in modeling speech-based lexicon compared to text and phone-based modalities, which is mainly caused by factor C. The representation of the same semantic unit, such as a word, is basically consistent in text and phone modalities. Recognizing valid words in these modalities is an empirical task, requiring only a judgment of whether the input has appeared in the training data. For speech LMs, however, the representation of the same text token or phonetic unit word would be combinatorial exploded. Lexical modeling in speech demands strong generalization capabilities, which are extremely challenging under unsupervised training. Since the positive examples in sWUGGY consist of infrequent words, the disadvantage of \texttt{Speech-HuBERT} is further amplified.

For the syntactic task, factor A still has a minor impact. Under the influence of factor B, the accuracy of \texttt{Phone-Repeat} decreased by 11.1\%. This suggests that adding the uncertainty of duration increases the difficulty of syntactic modeling. Furthermore, factor C introduced additional complexity through paralinguistic information, and the unsuccessful lexical modeling makes syntax recognition even harder. As a result, the accuracy of \texttt{Speech-HuBERT} drops by 13.4\% compared to \texttt{Phone-Repeat}.

For the semantic task, the accuracy of the LMs gradually decreases under the influence of factors B and C. It declines from 66.6\% in \texttt{Phone-Raw} to 58.3\% in \texttt{Phone-Repeat}, and finally to 52.9\% in \texttt{Speech-HuBERT}.

For the continuation task, both factors B and C significantly impact generation quality, with the perplexity increasing sharply. They bring 88.3\% and 140.7\% PPL increases, respectively. This highlights that duration variability and paralinguistic complexity severely challenge the language model's ability to maintain coherent and high-quality generation over extended sequences.

\subsection{Data Scaling Analysis}
Following the methodology of scaling laws~\cite{kaplan_scaling_2020}, this subsection compares LMs trained with an equivalent amount of computational resources. In this work, since we train models of the same size, we measure computational effort by the number of tokens the model has processed within the first epoch of training. For each objective task, we evaluate the LM checkpoints across various stages of progress within the first epoch. The results are presented in Figure \ref{fig:main-res-scale}. Each point in the figure corresponds to a specific checkpoint, where the x-axis represents the number of trained tokens, and the y-axis denotes the corresponding task accuracy. The points are color-coded to distinguish between different modalities. 


Almost all straight lines fit their respective point sets well, and except for the combination of (sWUGGY, \texttt{Speech-HuBERT}), all slopes are positive. 
It can be observed that, for lexical tasks, except for \texttt{Speech-HuBERT}, the $k$ values of other modality LMs are approximately similar, indicating that factor C has the most significant impact. This result aligns with Table \ref{tab:infl-factor-cvg}. Similarly, in syntactic tasks, both factors B and C negatively affect the scaling speed of the models. In semantic tasks, factors A, B, and C all influence performance scaling to some extent.

\subsection{Analysis on Internal Outputs}

We observe that speech LMs face particular challenges in learning the lexicon. While all modalities eventually converge to similar accuracies on sWUGGY with comparable scaling slopes, their lexical modeling trajectories within cascaded Transformer layers differ. To investigate, we analyze intermediate layers by projecting LM hidden states through the output layer to obtain multinomial distributions, which are treated as intermediate representations. Figure~\ref{fig:swuggy-by-layers} presents sWUGGY accuracies across layers for each modality. In early layers, \texttt{Text-BPE} and \texttt{Phone-BPE} learn lexical patterns most quickly, as BPE tokens inherently encode lexical priors. \texttt{Text-Raw} and \texttt{Phone-Raw} follow, since they require integrating multiple characters or phones to reach the word level. \texttt{Phone-Repeat} lags behind because duration-based repetition expands the lexical space, while speech tokens exacerbate the issue by creating a combinatorial explosion that prevents the model from consistently ``memorizing'' lexical units.


To illustrate, we further compare \texttt{Text-BPE}, \texttt{Phone-Raw}, and \texttt{Phone-Repeat} across tasks (Figures~\ref{fig:text-bpe-by-layers} and \ref{fig:phone-repeat-by-layers}). Despite achieving similar final sWUGGY accuracy, their intermediate behaviors diverge substantially. Lexical modeling emerges as the foundation for syntax and semantics: modalities that fail to stabilize lexical representations early struggle to acquire higher-level structure later. Semantic-dense modalities, such as \texttt{Text-BPE} and \texttt{Phone-BPE}, consistently map the same semantic unit (e.g., a word) to stable token sequences, enabling efficient lexical learning in shallow layers. By contrast, \texttt{Phone-Repeat} and speech modalities lack this consistency due to variable pronunciation and representation, which hampers reliable lexical grounding and, in turn, syntax and semantic modeling.


These experiments reveal why speech-based modalities are harder to train. Factor A (phonetic information) has only minor impact. Factor B (longer sequence length) increases difficulty by introducing duration variability, which complicates syntactic and semantic modeling. Factor C (paralinguistic information) adds another layer of variability, severely degrading lexical learning. Even when using discrete HuBERT tokens that reduce paralinguistic content, language modeling remains markedly more challenging than for text or phone-based modalities.

\section{Future Directions}
\label{sec:future-dir}

This study suggests that robust lexical-level modeling is a critical prerequisite for building effective end-to-end SLMs. To advance in this direction, two key issues—long sequence length (Factor B) and paralinguistic variability (Factor C)—must be addressed. We outline two promising directions:

\paragraph{Shortening Speech Sequence Length} 
Reducing sequence length remains a central challenge. Fixed-length solutions such as low-bit-rate codecs or uniform downsampling can reduce information rate, but often suffer from mismatches between frame boundaries and semantic units. Variable-length approaches, by contrast, appear more promising, as illustrated by the efficiency of \texttt{Phone-BPE}. However, designing a simple, variable-length, low-frame-rate representation that maintains high resynthesis quality is still an open problem.

\paragraph{Extra Semantic Supervision} 
Augmenting training with stronger lexical-level semantic supervision may further improve SLMs. Existing strategies, such as data interleaving or reinforcement learning, provide only weak and indirect signals, leading to limited gains. More explicit supervision—e.g., time-aligned lexical or semantic annotations—could help models establish consistent lexical grounding, thereby enhancing both training effectiveness and final performance.

\section{Conclusion} 
We conducted a systematic analysis of performance degradation from text LMs to speech LMs, and identified the major factors impeding speech LMs. Among them, paralinguistic variability (Factor C) exerts the strongest influence, especially on lexical modeling, while longer sequence length (Factor B) also poses significant difficulty. These findings underscore the importance of lexical-level modeling as the foundation for higher-level semantics. Based on this insight, we highlight future directions focused on shortening speech sequences and incorporating stronger semantic supervision, which may help bridge the gap between speech LMs and text LLMs.

\section{Acknowledgement}
This work has been supported by the China NSFC Project (No. 92370206) and the Key Research and Development Program of Jiangsu Province, China (No.BE2022059).


\bibliographystyle{IEEEtran}
\bibliography{main}

\end{document}